\documentclass[fleqn,usenatbib]{mnras}

\usepackage{newtxtext,newtxmath}
\usepackage{textgreek}

\usepackage[T1]{fontenc}
\usepackage[utf8]{inputenc}
\newcommand{\mycomment}[1]{}
\newcommand{\edit}[1]{\textcolor{black}{#1}}
\newcommand{\newedit}[1]{\textcolor{black}{#1}}

\DeclareRobustCommand{\VAN}[3]{#2}
\let\VANthebibliography\thebibliography
\def\thebibliography{\DeclareRobustCommand{\VAN}[3]{##3}\VANthebibliography}

\usepackage{graphicx}	
\usepackage{amsmath}	

\title[Atmospheric Refraction in Fireball Analysis]{Distance-Independent Atmospheric Refraction Correction for Accurate Retrieval of Fireball Trajectories}

\author[J. Visuri et al.]{
Jaakko Visuri,$^{1}$
Maria Gritsevich, $^{2,1,3}$\thanks{E-mail: maria.gritsevich@helsinki.fi}
Janne Sievinen$^{1}$
\\
$^{1}$Finnish Fireball Network, Ursa Astronomical Association, Finland\\
$^{2}$Faculty of Science, Gustav Hällströmin katu 2, FI-00014 University of Helsinki, Finland\\
$^{3}$Institute of Physics and Technology, Ural Federal University\\
}

\pubyear{2026}

\begin{document}
\label{firstpage}
\pagerange{\pageref{firstpage}--\pageref{lastpage}}
\maketitle

\begin{abstract}
Accurate determination of fireball direction is essential for retrieving trajectories and velocities. Errors in these measurements have significant implications, affecting the calculated pre-impact orbit, influencing mass estimates, and impacting the accuracy of dark flight simulations, where applicable. Here we implement a new atmospheric refraction correction technique that addresses a significant aspect previously overlooked in the field of meteor science. Traditional refraction correction techniques, originally designed for objects positioned at infinite distances, tend to overcompensate when applied to objects within the Earth's atmosphere. To rectify this issue, our study introduces the concept of the atmospheric refraction δz-correction technique (hereafter δz-correction technique), involving the artificial elevation of the observer's site height above sea level. We utilize analytically derived formulas for the δz-correction in conjunction with commonly used refraction models, validating these results against a numerical solution that traces light rays through the atmosphere. This meticulous ray-tracing model is applied to finely meshed atmospheric layers, yielding precise correction values. \edit{We evaluate multiple sources of error in order to quantify the achievable accuracy of the proposed method. Our approach (1) enables the determination of fireball positions with improved astrometric accuracy, (2) removes the explicit dependence on the fireball’s distance from the observer or its height above Earth’s surface within the limits imposed by realistic atmospheric variability, and (3) simplifies meteor data processing by providing a robust framework for analysing low-elevation fireball observations, for which atmospheric refraction is significant and is automatically corrected by the method.
As a result of this work, we provide an open, publicly accessible software for calculating the \textdelta z-correction.}
\end{abstract}




\section{Introduction}
Meteorites, as celestial messengers from the far reaches of our solar system, hold invaluable clues about the universe's composition and history. To study these valuable specimens and gain insights into their origins, accurate measurements of their entry trajectories and velocities are essential (Ceplecha 1987; Trigo-Rodríguez et al. 2015; Dmitriev et al. 2015; Meier et al. 2017; Vida et al. 2018; Jansen-Sturgeon et al. 2019; Kyrylenko et al. 2023; Peña-Asensio et al. 2025a; Jenniskens and Devillepoix 2025; Shober et al. 2026). Deriving accurate fireball directions is a complex task, riddled with challenges that can lead to inaccuracies (Borovicka et al. 1995; Egal et al. 2017; Sansom et al. 2021).

The ramifications of inaccurate direction measurements extend beyond mere scientific curiosity; they have profound implications for understanding not only the pre-impact orbits including identification of potential hypervelocity impactors (Peña-Asensio et al. 2024), but also the simulations of their dark flight through Earth's atmosphere (Vinnikov et al. 2016; Moilanen et al. 2021; Towner et al. 2022). Indeed, these inaccuracies propagate, influencing the calculated strewn field area where survived meteorite fragments could be recovered (Wetherill and ReVelle 1981, Gardiol et al. 2021, Andrade et al. 2023). \edit{Modern tools, including the \textalpha -\textbeta -model (Gritsevich 2007, 2008, 2009) and Bayesian filtering techniques (Sansom et al. 2015, 2017), estimate the complete dynamical state of a meteoroid along its path, but remain sensitive to uncertainties in trajectory direction and velocity.}

One of the factors contributing to these inaccuracies is atmospheric refraction, a phenomenon in which light or radio waves passing through Earth's atmosphere are bent due to variations in air density with altitude. This bending of waves may affect the apparent position of objects, including satellites, when viewed from the Earth's surface (Green 1985; Saemundsson 1986, Stone 1996; Auer et al. 2000). Atmospheric refraction must be taken into account to accurately determine the object positions in the sky.

When a fireball appears at a low elevation angle, applying the full refraction correction, \edit{assuming the object is at infinity}, can result in an overly large adjustment (Visuri et al. 2020, 2021). \edit{Part of the refraction, when calculated along the line of sight assuming a larger distance, occurs in the upper atmosphere at altitudes above the fireball itself.} However, it is essential to recognize that this effect is, for practical purposes, negligible.

A more substantial source of error arises from the curvature of the trajectory between the fireball and the observing station. Correcting the elevation angle with the full refraction in such cases can lead to a fireball position significantly below its actual location. This effect, although well acknowledged in other fields such as satellite tracking, has not been implemented in processing fireball observations until this work. This oversight is remarkable considering that the meteorite-producing fireballs often traverse at only a few tens of kilometers above the Earth's surface.

This study bridges this gap by introducing a refraction correction technique that enhances fireball data analysis. We demonstrate how more accurate fireball directions \edit{and velocities can be retrieved using the proposed \textdelta z-correction and examine its mathematical basis, practical implementation, and physical implications. By enabling a more robust determination of fireball positions, the method improves meteorite fall predictions for current observations and allows historical fireball events to be re-analysed with higher accuracy, providing a more reliable foundation for revisiting past cases and refining existing datasets (Castro-Tirado et al. 2008; Gritsevich et al. 2024).}

\section{ \textdelta z-correction technique}
The atmospheric refraction effect is relatively well studied and accounted for when correcting for satellite parallax (Green 1985; Tausworthe 2005). By implementing this correction, the precision of satellite tracking and navigation is significantly enhanced. This adjustment accounts for the bending of light as it passes through the Earth's atmosphere, thereby allowing for accurate determination and alignment of the satellite apparent position with its actual position in the celestial sphere. This correction is imperative for various applications, including satellite-based communication, navigation systems, and astronomical observations, where precise knowledge of the satellite's location is paramount.

\begin{figure}
\includegraphics[width=\columnwidth]{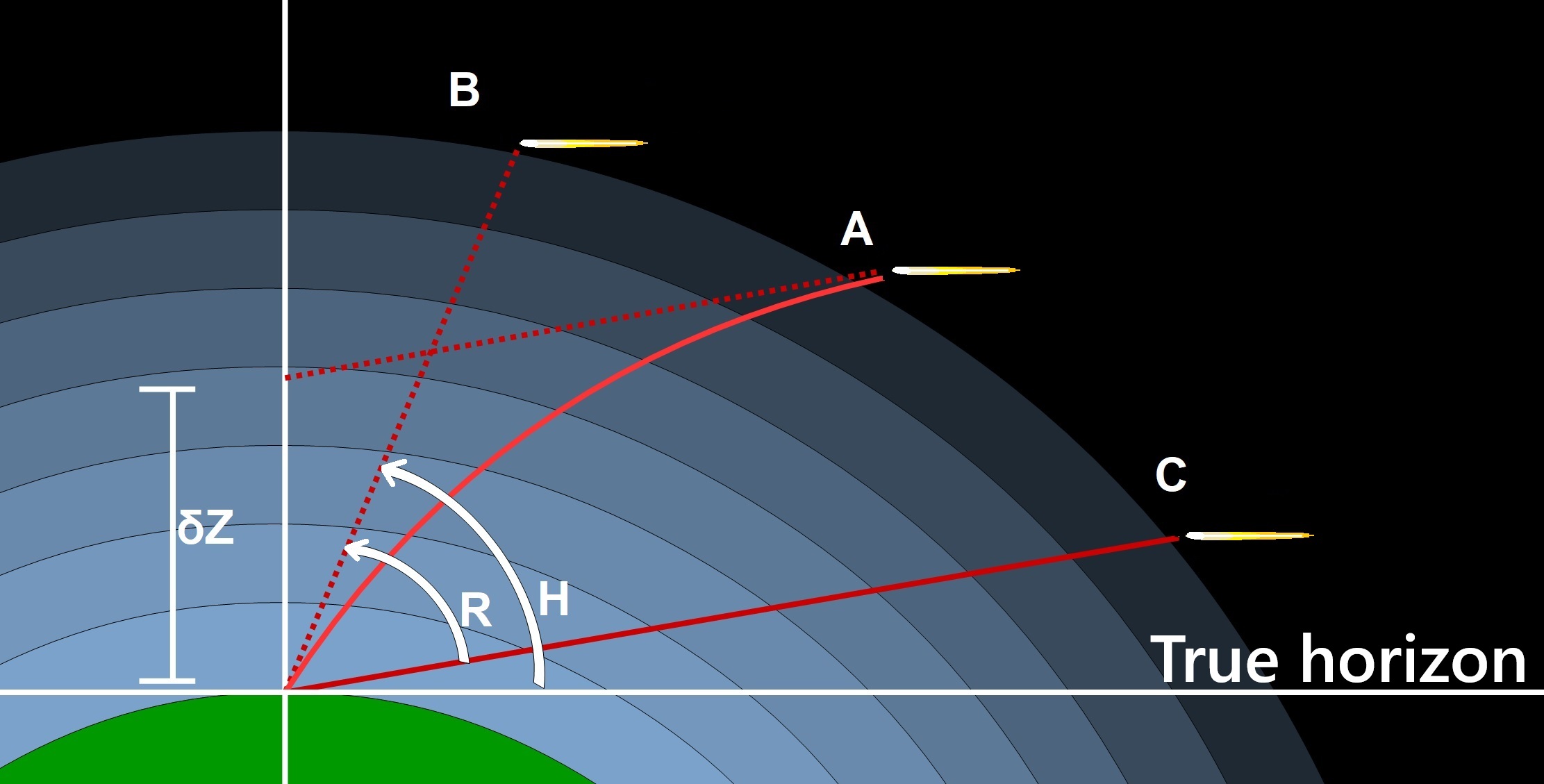}
\caption{ \edit{Geometry of the \textdelta z correction for a fireball observation. 
A fireball located at point A is seen at the apparent position B due to atmospheric refraction. 
Applying the standard refraction correction for an object at infinite distance shifts the direction to point C, which is appropriate for stars but not for fireballs observed at finite range. 
To make the refraction-corrected direction intersect the true fireball position, the observing site must be virtually raised by an amount \textdelta z, as illustrated. 
The apparent elevation angle $H$ is positive above the true horizon and negative below it; the \textdelta z\ correction is always positive upward, and the refraction $R$ is defined as positive in the upward direction.}}
\label{fig:intro}
\end{figure}

Figure 1 illustrates a \edit{typical fireball observation scenario. The fireball position in the atmosphere is at point A. Due to atmospheric refraction, it is observed at point B. Applying the standard refraction correction moves the apparent position to point C, which would be accurate for distant stars but does not correspond to the true position of a nearby fireball. To align the refraction-corrected direction with the actual position of the fireball, the observation station must be artificially elevated by an amount \textdelta z, as indicated in Figure 1.}

This approach provides a practical, distance-independent correction for low-elevation fireball observations: with the  \textdelta z adjustment, the direction computed using the full refraction correction aligns with the true position of the fireball (Lyytinen E., personal communication). Consequently, this \textdelta z value can be derived into an analytical formula for a spherically symmetric atmosphere (Green 1985):

\begin{equation}
\label{eq:green}
\delta z = r_0 \dfrac{n_0 sin(90-H)}{sin(90-H + R) - 1}
\end{equation}

where H is the apparent elevation angle and R is the full refraction corresponding to H. \edit{Both H and R are expressed in degrees.} The refractive index at sea level is n\textsubscript{0} and r\textsubscript{0} refers to the Earth's radius at the observing site, \edit{that is, the Earth’s mean radius plus the height of the observer}.
The Green model is highly dependent \edit{on the refraction formulation adopted and is therefore particularly sensitive to errors at lower apparent elevations, as illustrated in the last column of Table 1. We compare this analytical formulation in parallel with the developed numerical \textdelta z-correction method described in section 3.}

\edit{In this work, the apparent elevation $H$ is defined with respect to the geometric (true) horizon: 
$H>0^\circ$ denotes objects above the geometric horizon and $H<0^\circ$ objects below it. 
The refraction $R$ is defined as positive upward, such that the apparent elevation is
\[H = H_{\rm true} + R .
\]
The $\delta z$ correction is defined as a positive upward vertical shift of the observing site.}

\edit{For negative apparent elevations ($H<0^\circ$), empirical refraction formulae such as Saemundsson (1986) and Bennett’s model discussed below
are not formally validated and become  increasingly unreliable close to and below the horizon, 
where refraction gradients are strongest. 
In these cases we do not extrapolate the analytical formulae. 
Instead, $\delta z$ is computed exclusively by a ray-tracing model through the layered atmosphere (see section 3), 
which remains well defined for both positive and negative $H$.}

One commonly used refraction model is Bennett's model (Bennett 1982):

\begin{equation}
\label{eq:bennet}
R = \dfrac{(0.28*P / (T+273.15))}{(tan(H + 7.31/(H+4.4)))} / 60
\end{equation}

Here, T represents the temperature in degrees Celsius, and P denotes the pressure in millibars at the observing site. \edit{Both R and H are expressed in degrees.} The model employed in Equation \ref{eq:green} provides \textdelta z corrections exclusively when Bennett's refraction model is deemed valid.

It is crucial to note that the Green model yields a \textdelta z value for a light ray that traverses the entire atmosphere. However, phenomena such as meteors can occur at various heights within the atmosphere, rendering traditional refraction corrections inadequate. Using the full refraction correction independently or \edit{the Green model itself for such meteors} would result in an excessive compensation of the fireball position.

To study the extent of true refraction and the necessary \textdelta z correction, the atmosphere must be segmented into distinct layers, with the path of light rays integrated within each layer. It is well-established that the majority of refraction phenomena occur within the lower, denser layers of the atmosphere. \edit{At higher altitudes, where density gradients are weak, the light path can be approximated as nearly straight compared to the strongly curved ray segment in the lower atmosphere.} Consequently, the full \textdelta z value remains reasonably accurate, to some extent, even for bolides that penetrate deeper in the atmosphere.

Our primary objective is to investigate the error associated with the geometrical position when we apply the \textdelta z value derived using the full refraction model, as illustrated in Figure 2. A secondary goal is to explore applicability of alternative refraction models and corrections to the refraction values employed in the Green model.

\section{Refraction using the ray-tracing model}

Numerical data were generated through a ray-tracing \newedit{(hereafter RT)} model used to simulate the way light bends when propagating in the atmosphere. The atmosphere was discretized into concentric spherical shells, each 10 meters thick, spanning from mean sea level (at height = 0 m) to an upper bound of 86,000 meters. Negligible contributions from higher atmospheric layers were excluded.

For each shell, air pressure, temperature, and density values were derived from the \edit{U.S. Standard Atmosphere (USSA) model (NASA, 1976)}, that is generally recommended for meteor studies (Lyytinen and Gritsevich, 2016). \edit{The USSA provides a physically consistent reference profile for pressure and temperature as a function of altitude above sea level when site-specific atmospheric data are unavailable}.

The RT model was conducted in reverse order. In reality, light \edit{propagates from the emitting object through the atmosphere and reaches the observer at a given apparent angle. In this study, however, rays were initiated at the observer and traced upward through successive atmospheric layers. Within each layer, the ray was assumed to follow a straight path, with refraction applied at each layer boundary according to} Snell's law:
\begin{equation}
n_1 * sin(\delta_1) = n_2 * sin(\delta_2)
\end{equation}
where n and \edit{\textdelta}  are the refractive indices and wavefront angles respectively. 
After undergoing refraction at each layer boundary, the light ray continued to propagate in a straight line until reaching the upper boundary. The refractive index at sea level, n\textsubscript{0}, was calculated using the complex Ciddor equation (Ciddor 1996), with input parameters including a vacuum wavelength of 570 nm, a CO\textsubscript{2} concentration of 0 ppm, a temperature of 15 °C, an atmospheric pressure of 1013.25 mbar, and a relative humidity of 80 \%. Based on these conditions, the resulting refractive constant for air \edit{was determined to be n\textsubscript{air} =0.000276567}.

The refractive indices for the atmospheric layers were then computed using the following equation:
\begin{equation}
n_i = 1 + n_{air} * (P_i / P_0) * (T_0 / T_i)
\end{equation}
where P\textsubscript{i} and T\textsubscript{i} represent the air pressure and temperature at the i-th layer 
and P\textsubscript{0} and T\textsubscript{0} denote the sea level air pressure and temperature, respectively. The values for pressure and temperature were derived from the U.S. Standard Atmosphere (USSA) model. \edit{As a result, we obtained the refractive index for sea level n\textsubscript{0} = 1.000276567}.
Changes in air humidity and carbon dioxide content were disregarded due to their negligible impact on the results.

\edit{Beyond the upper integration limit, the light ray was treated as an incoming ray and propagated back towards the lower atmosphere along a straight path, without further atmospheric refraction. At the observer’s location, the vertical offset between sea level and this unrefracted} path was calculated and designated as \textdelta z. This offset is illustrated in Fig. 1 by the dotted line extending from point A.

In order to validate the accuracy of our RT model, we conducted a comparative analysis with the Green model (equation \ref{eq:green}) and 
Bennett's refraction (equation \ref{eq:bennet}).
To achieve this, a light ray was emitted at 21 different angles ranging from 0 to 20 degrees with a 1-degree interval. For each of these angles, the RT model was employed to determine total refraction and \textdelta z values. In Table 1, we present the apparent angles, total refraction, and
\textdelta z values obtained through our RT model. Additionally, we provide corresponding values derived from equations (\ref{eq:green}) and (\ref{eq:bennet}), along with the differences in comparison to our RT model. The successful validation of the RT model is evident, particularly through the equivalence observed between the Green model and our \textdelta z values, as shown in Table 1. This emphasizes the reliability and accuracy of the applied RT methodology.

The horizon appears to lower at an elevation of 1,000 meters above the mean sea level (m.a.s.l.) due to both Earth's spherical shape and atmospheric refraction. According to our RT model, a ground level is observed at an apparent angle of H = -0.92807 \textdegree. This implies the possibility of observing fireballs at negative apparent angles.

To investigate this phenomenon, the RT model was applied for angles ranging from -0.1 to -0.9, utilizing 2 meters thick atmospheric layers to ensure sufficient accuracy. The results are detailed in table \ref{tab:neg}. Additionally, \textdelta z values for positive apparent angles are compared to their corresponding values at sea level in the same table. This comparison demonstrates the impact of atmospheric conditions on observations at different altitudes.

\edit{We can demonstrate that numerical accuracy and convergence are satisfactory when using fixed atmospheric layer thicknesses of 10 m, and 2 m for low-elevation cases, without the need for adaptive layering. The convergence behaviour is summarised in Appendix (Table \ref{tab:conv}). Adopting a fixed layer thickness ensures fast computation while avoiding the repeated step-size adjustments and conditional checks required by adaptive integration schemes.} 

\edit{For negative apparent elevation angles, evaluated here for an observing site at 1000 m a.s.l., the choice of layer thickness becomes more significant. The difference between using 0.01 m and 2 m layers amounts to 14.75 m at an apparent elevation of -0.5° and 49.29 m at -0.9°. Nevertheless, a 2 m layer thickness is sufficient for most practical applications, as observations of fireballs at negative apparent elevations are extremely rare.}

\begin{table*}
\caption{Errors for \textdelta z value when object is situated lower in the atmosphere (height le; 30,000 meters). 
\textdelta z errors are shown for five different heights. 
The error is the vertical difference of the true location of the object compared to the full \textdelta z. 
This is illustrated in Fig 3.}
\begin{tabular}{ c | c | c c c c c}
\hline
Apparent & Full  &  \multicolumn{5}{c}{Object height} \\
angle & \textdelta z & 10,000 m& 15,000 m& 20,000 m& 25,000 m& 30,000 m\\\
[\textdegree] & [m] & [m] & [m] & [m] & [m] & [m] \\
\hline
0 & 2040.429 & 124.235 & 42.395 & 15.374 & 5.853 & 2.331 \\
1 & 1153.166 & 116.242 & 40.389 & 14.798 & 5.672 & 2.270 \\
2 & 701.180 & 97.727 & 35.410 & 13.313 & 5.193 & 2.104 \\
3 & 456.437 & 77.624 & 29.447 & 11.421 & 4.556 & 1.877 \\
4 & 314.514 & 60.553 & 23.894 & 9.542 & 3.892 & 1.632 \\
5 & 227.109 & 47.370 & 19.275 & 7.888 & 3.282 & 1.399 \\
\hline
\end{tabular}
\label{tab:conv}
\end{table*}

\begin{figure}
\includegraphics[width=\columnwidth]{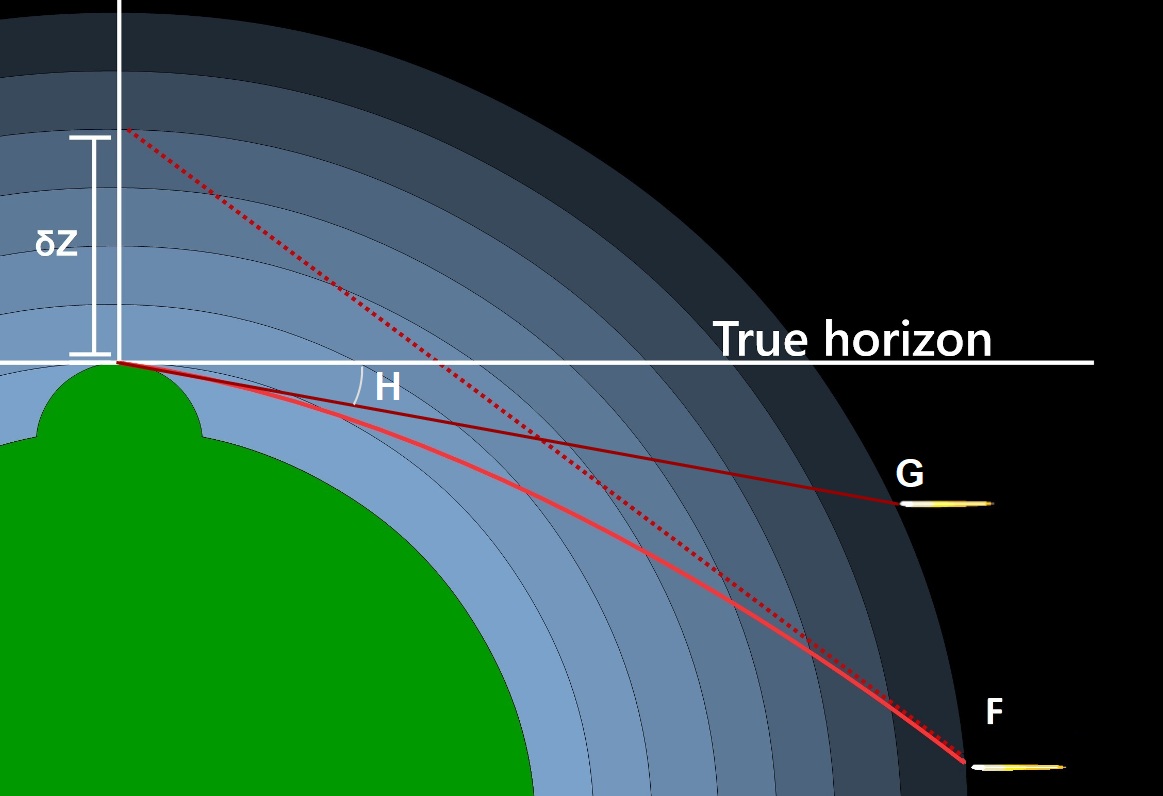}
\caption{Negative apparent elevation angles may occur for observers at elevated sites. \newedit{The fireball is observed at an apparent angle H, appearing to be at point G. Due to atmospheric refraction, the fireball is actually located at point F. The \textdelta z correction is applied to adjust the refraction-corrected line of sight so that it aligns with the true position of the fireball.}
}
\label{fig:neg}
\end{figure}

\begin{table*}
\caption{Refraction and \textdelta z values for observing site at 0 m.a.s.l. The refractions calculated by our RT model for the corresponding apparent angles are compared with those computed using Bennett's equation, along with the resulting differences.
\textdelta z values for total refraction in the third last column corresponds to the \textdelta z shown in Fig 1.
\textdelta z values in the last two columns correspond to the \edit{equation (1)}. In the second-to-last column, refractions are provided using the RT model, and in the last column, refractions are determined by Bennett's equation (3).}
\begin{tabular}{ c | c c c | c c c }
    \hline
    \multicolumn{1}{p{2cm}|}{\begin{center}Apparent \newline angle \end{center}} & 
    \multicolumn{1}{p{2cm}}{\begin{center}Total refraction \newline as RT model \end{center}} & 
    \multicolumn{1}{p{2cm}}{\begin{center}Total refraction \newline as Bennett's model \end{center}} & 
    \multicolumn{1}{p{2cm}}{\begin{center}Difference \newline in refraction \newline (RT - Bennett's)\end{center}} & 
    \multicolumn{1}{p{2cm}}{\begin{center}\textdelta z \newline as RT model\end{center}} & 
    \multicolumn{1}{p{2cm}}{\begin{center}\textdelta z \newline as Green model, \newline \edit{refraction adopted using RT}\end{center}} & 
    \multicolumn{1}{p{2cm}}{\begin{center}\textdelta z \newline as Green model, \newline \edit{refraction adopted using}\newline Bennett's\end{center}} \\\
    [\textdegree] & [\textdegree] & [\textdegree] & [\textdegree] & [m] & [m] & [m] \\
    \hline
    0 & 0.5334 & 0.5658 & -0.0324 & 2040 & 2040 & 2075 \\
    1 & 0.3906 & 0.3992 & -0.0087 & 1153 & 1153 & 1143 \\
    2 & 0.2951 & 0.2989 & -0.0038 & 701 & 701 & 688 \\
    3 & 0.2332 & 0.2354 & -0.0022 & 456 & 456 & 444 \\
    4 & 0.1907 & 0.1926 & -0.0019 & 315 & 315 & 300 \\
    5 & 0.1604 & 0.1622 & -0.0018 & 227 & 227 & 209 \\
    6 & 0.1378 & 0.1396 & -0.0019 & 170 & 170 & 148 \\
    7 & 0.1204 & 0.1223 & -0.0019 & 132 & 132 & 106 \\
    8 & 0.1068 & 0.1086 & -0.0019 & 105 & 105 & 75 \\
    9 & 0.0957 & 0.0976 & -0.0019 & 85 & 85 & 52 \\
    10 & 0.0866 & 0.0885 & -0.0018 & 71 & 71 & 34 \\
    11 & 0.0791 & 0.0808 & -0.0018 & 59 & 59 & 20 \\
    12 & 0.0726 & 0.0744 & -0.0017 & 51 & 51 & 9 \\
    13 & 0.0671 & 0.0688 & -0.0017 & 44 & 44 & 0 \\
    14 & 0.0623 & 0.0639 & -0.0016 & 38 & 38 & -7 \\
    15 & 0.0581 & 0.0597 & -0.0015 & 33 & 33 & -13 \\
    16 & 0.0544 & 0.0559 & -0.0015 & 30 & 30 & -18 \\
    17 & 0.0511 & 0.0526 & -0.0014 & 26 & 26 & -23 \\
    18 & 0.0482 & 0.0495 & -0.0014 & 24 & 24 & -27 \\
    19 & 0.0455 & 0.0468 & -0.0013 & 22 & 22 & -30 \\
    20 & 0.0431 & 0.0444 & -0.0013 & 20 & 20 & -32 \\
    \hline
    \hline
\end{tabular}
\label{tab:base}
\end{table*}

\begin{table*}
\caption{Refraction and \textdelta z  values for observing site at 1,000 m.a.s.l. 
The second column shows the full refraction (for a star) observed at the corresponding apparent angle. 
Differences relative to the \textdelta z values for sea level are shown in the fourth column. 
The fifth column shows \textdelta values calculated using the Green model with Bennett's refraction, incorporating corresponding pressure and temperature values from the USSA model. The atmosphere was discretized into 2-meter thick layers for negative angles. All values are rounded to four decimal places.}
\begin{tabular}{ c | c c c c }
\hline
\multicolumn{1}{p{2cm}}{\begin{center}Apparent angle\end{center}} & 
\multicolumn{1}{p{2cm}}{\begin{center}Total refraction \newline RT-model\end{center}} & 
\multicolumn{1}{p{2cm}}{\begin{center} \textdelta z \newline RT-model \end{center}} & 
\multicolumn{1}{p{2cm}}{\begin{center} \textdelta z difference. \newline (sea level - 1000m) \end{center}} &
\multicolumn{1}{p{2cm}}{\begin{center} \textdelta z \newline Green model \newline (Bennett's refraction with USSA values) \end{center}} \\\
[\textdegree] & [\textdegree] & [m] & [m] & [m] \\
\hline
-0.9 & 0.7049 & 3318 & - & 3521\\
-0.8 & 0.6756 & 3095 & - & 3309\\
-0.7 & 0.6490 & 2894 & - & 3109\\ 
-0.6 & 0.6296 & 2721 & - & 2919\\ 
-0.5 & 0.5980 & 2530 & - & 2742\\ 
-0.4 & 0.5748 & 2369 & - & 2576\\ 
-0.3 & 0.5463 & 2210 & - & 2421\\ 
-0.2 & 0.5383 & 2092 & - & 2277\\ 
-0.1 & 0.5081 & 1951 & - & 2144\\ 
0 & 0.4867 & 1831 & 209 & 2020\\
1 & 0.3560 & 1032 & 121 & 1187\\ 
2 & 0.2687 & 626 & 75 & 781\\ 
3 & 0.2121 & 407 & 50 & 562\\ 
4 & 0.1734 & 280 & 35 & 433\\ 
5 & 0.1457 & 202 & 25 & 352\\ 
6 & 0.1252 & 151 & 19 & 297\\ 
7 & 0.1094 & 117 & 15 & 259\\
8 & 0.0969 & 93 & 12 & 231\\ 
9 & 0.0869 & 76 & 10 & 210\\
10 & 0.0787 & 67 & 8 & 194\\ 
\hline
\end{tabular}
\label{tab:neg}
\end{table*}

\section{Error Estimation}
\subsection{Meteor height}
The height of the fireball above the ground level influences the error associated with the full refraction correction. A fireball at a lower height from the sea level necessitates a smaller correction to the \textdelta z value. This discrepancy, represented by the error in \textdelta z, is visually depicted in Figure 3. The RT model was employed to investigate the error of the \textdelta z for objects situated in the lower atmospheric layers. We examined \textdelta z errors for objects positioned at 10,000, 15,000, 20,000, 25,000, and 30,000 meters above sea level. The corresponding error values are detailed in the Table \ref{tab:lower}.

\begin{figure}
{\includegraphics[width=\columnwidth]{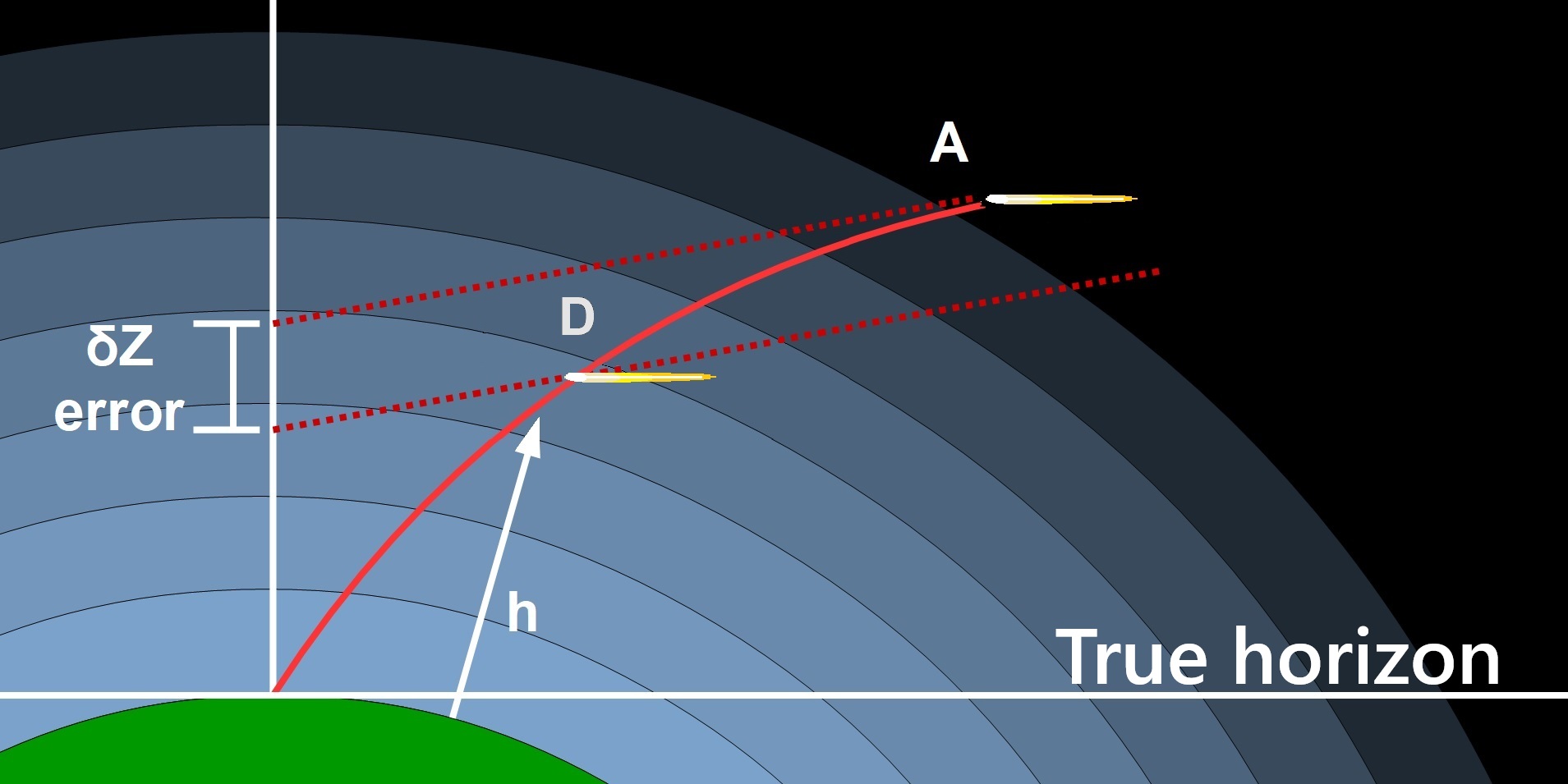}
\caption{\newedit{The fireball might be located at a lower atmospheric altitude at point D. If the full \textdelta z correction is applied, the refraction-corrected line of sight aligns perfectly with point A. The difference between this line of sight and point D represents the error in \textdelta z.}}}
\label{fig:error}
\end{figure}

\begin{table*}
\caption{Errors for \textdelta z value when object is situated lower in the atmosphere (height < 30,000 meters). 
\textdelta z errors are shown for five different heights. 
The error is the vertical difference of the true location of the object compared to the full \textdelta z. 
This is illustrated in Fig 3.}
\begin{tabular}{ c | c | c c c c c}
\hline
Apparent & Full  &  \multicolumn{5}{c}{Object height} \\
angle & \textdelta z & 10000 m& 15000 m& 20000 m& 25000 m& 30000 m\\\
[\textdegree] & [m] & [m] & [m] & [m] & [m] & [m] \\
\hline
0 & 2040.429 & 124.235 & 42.395 & 15.374 & 5.853 & 2.331 \\
1 & 1153.166 & 116.242 & 40.389 & 14.798 & 5.672 & 2.270 \\
2 & 701.180 & 97.727 & 35.410 & 13.313 & 5.193 & 2.104 \\
3 & 456.437 & 77.624 & 29.447 & 11.421 & 4.556 & 1.877 \\
4 & 314.514 & 60.553 & 23.894 & 9.542 & 3.892 & 1.632 \\
5 & 227.109 & 47.370 & 19.275 & 7.888 & 3.282 & 1.399 \\
\hline
\end{tabular}
\label{tab:lower}
\end{table*}

\subsection{Atmospheric anomalies}
 Understanding how atmospheric anomalies affect refraction is crucial for accurate weather predictions, astronomical observations, and various communication technologies. Atmospheric anomalies wield an influence, causing deviations in the paths of light or sound waves as they traverse distinct layers of the atmosphere. Temperature variations, a prevalent form of atmospheric anomaly, contribute to shaping refraction patterns. During heatwaves characterized by steep temperature gradients, the air density undergoes substantial variations, leading to the pronounced bending of light rays. This bending can manifest as mirages, causing distant objects to appear distorted or displaced from their actual positions. Conversely, in cold snaps, the denser, colder air near the surface can induce downward bending of light, extending the range of visibility.

Therefore we explored the influence of thermal inversion on \textdelta z values. We simulated a scenario where the inversion layer is positioned 200 meters above ground level, a common altitude for thermal inversions in suburban areas and mountain valleys. Our results, detailed in \ref{tab:inv} are compared to the values presented in table \ref{tab:base}. 

The simulation of the inversion layer at 200 meters above ground (sea) level involved reducing the temperatures from the USSA model by 5 degrees. This adjustment created an inversion layer with a temperature difference \textDelta T of 5 degrees at the specified height.

\subsection{\edit{Real atmospheric data}}

\edit{The U.S. Standard Atmosphere represents a climatological mean and does not capture short-term or site-specific atmospheric variability. When available, the use of real atmospheric data can therefore improve the accuracy of δz estimates. In our showcase event FN200907, the surface pressure at the observing site was 1008.5 mbar at 0 m a.s.l. This deviation from standard conditions contributes more significantly to the uncertainty in δz than numerical discretisation or convergence effects discussed in Section 3. As demonstrated in earlier studies, the use of standardised atmospheric models can constitute a major source of error in practical applications (Lyytinen and Gritsevich 2016).}

\subsection{\edit{Parameters of the Ciddor refractive index formulation}}


\edit{The refractive index in our RT model calculations is evaluated using the Ciddor equation, which allows the inclusion of wavelength, relative humidity, and adopting a realistic atmospheric CO$_2$ content of 385\,ppm (Abshire et al. 2010). Varying the wavelength across the visible range results in a refractive index change of approximately $3\times10^{-6}$, corresponding to a \textdelta z difference of about 12\,m. The effects of relative humidity and CO$_2$ concentration are even smaller, producing refractive index changes of order $6\times10^{-7}$ and $4\times10^{-8}$, respectively. These contributions are negligible compared to uncertainties arising from atmospheric pressure and temperature variability and can safely be ignored for broadband optical fireball observations.}

\subsection{\edit{Earth shape model}}


\edit{Both the Green model and our RT model assume a spherical Earth with concentric atmospheric shells. Adopting an ellipsoidal Earth model would introduce small latitude-dependent differences due to variations in the Earth’s radius. To quantify this effect, we evaluated \textdelta z for an observer at sea level under standard atmospheric conditions and an apparent elevation of $1^\circ$.}

\edit{Using an ellipsoidal Earth model yields $\textdelta z = 1152.88$\,m at the North Pole and $\textdelta z = 1149.64$\,m at the Equator, corresponding to a difference of approximately 3.2\,m. This effect is negligible compared to other sources of uncertainty and does not tangibly affect fireball trajectory reconstruction at the precision considered here.}

\begin{table*}
\caption{Comparison of refraction and \textdelta z in case of atmospheric inversion. Fireball is positioned at the height of 86,000 m. Table shows refraction and \textdelta z values using RT method described in this study. Values are presented for both scenarios, with and without inversion, allowing for direct comparison.}
\begin{tabular}{ c | c c | c c }
\hline
\multicolumn{1}{p{1.5cm}}{\begin{center}Apparent \newline angle \end{center}} & 
\multicolumn{1}{p{1.5cm}}{\begin{center}Total refraction \newline without inversion \end{center}} & 
\multicolumn{1}{p{1.5cm}}{\begin{center}Total refraction \newline with inversion \end{center}} & 
\multicolumn{1}{p{1.5cm}}{\begin{center}\textdelta z \newline without inversion\end{center}} & 
\multicolumn{1}{p{1.5cm}}{\begin{center}\textdelta z \newline with inversion \end{center}} \\\
[\textdegree] & [\textdegree] & [\textdegree] & [m] & [m] \\
\hline
0 & 0.5334 & 0.5797 & 2040 & 2122 \\
1 & 0.3906 & 0.4071 & 1153 & 1165 \\
2 & 0.2951 & 0.3036 & 701 & 704 \\
3 & 0.2332 & 0.2387 & 456 & 458 \\
4 & 0.1907 & 0.1949 & 315 & 315 \\
5 & 0.1604 & 0.1636 & 227 & 227 \\
\hline
\end{tabular}
\label{tab:inv}
\end{table*}

\section{\edit{Proof of Concept}}

\subsection{Showcase event FN200907}

We demonstrate the impact of the \textdelta z correction using a prominent example — an exceptionally durable fireball observed in Northern Scandinavia. This remarkable event occurred as the fireball entered the atmosphere on 2020-09-07 at 20:03:34 UTC, remaining visible for 26 seconds. In addition to Finnish Fireball Network (FFN) data, visual observations of the fireball were reported by 36 observers across Finland to the Ursa's Taivaanvahti ("SkyWarden") service https://www.taivaanvahti.fi. The locations of the fireball and observation stations are illustrated on the map in Fig 4.

\begin{figure}
\includegraphics[width=\columnwidth]{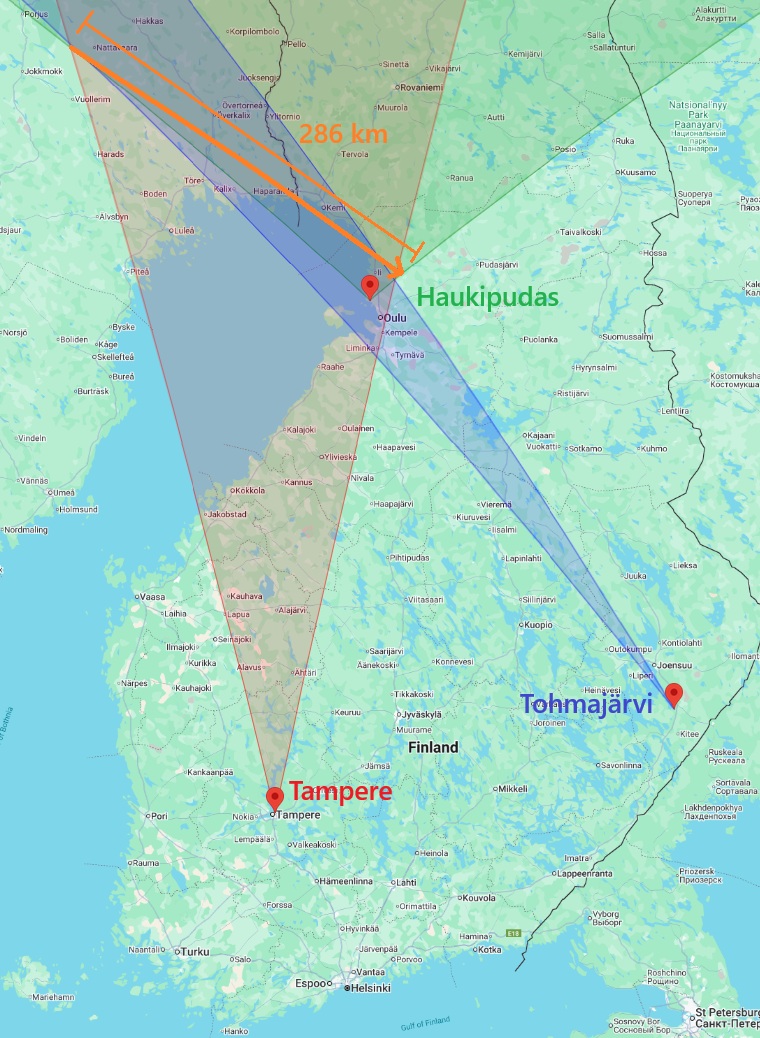}
\caption{Map of the FN200907 event showing Finnish Fireball Network stations used in the analysis (red markers). The fireball trajectory is shown as an arrow, originating from the northwest and moving towards the southeast.}
\label{fig:map}
\end{figure}

Image observations from two stations and one video observation were used to determine the luminous flight trajectory (see Appendix). Camera calibrations and fireball directions were determined using the FireOwl software by FFN (Visuri and Gritsevich 2021), a tool that has already demonstrated successful implementation in previous studies (Kyrylenko et al. 2023; Gritsevich et al. 2024; Peña-Asensio et al. 2024; Peña-Asensio et al. 2025b; Borderes-Motta et al. 2026). The measured values are presented in Table \ref{tab:setup} \edit{and are calculated using the WGS84 standard}.

\edit{At the Haukipudas station, which is close to the ground projection of the meteor, the apex of the luminous trajectory appeared approximately 58° above the horizon. Here, the δz-correction displaces the fireball plane by only 6 meters and can therefore be neglected}. Consequently, the Haukipudas observation, along with its corresponding directional data, serves as a reference plane for subsequent triangulation.

At the Tampere and Tohmajärvi stations, the fireball was observed nearly parallel to the horizon, appearing only a few degrees above it. This observational geometry permits the artificial elevation of each site, eliminating the need to apply separate \textdelta z corrections for both the beginning and end points of the trajectory. Instead, mean \textdelta z values are used, 244 m for Tohmajärvi and 248.5 m for Tampere. 

With these corrections applied, the adjusted directional data from Tampere and Tohmajärvi can be triangulated relative to the fireball plane defined by the Haukipudas observation. This enables the construction of distinct plane sections for each observing station. To evaluate the influence of the \textdelta z correction, results from triangulation both with and without the correction are compared. \edit{The corresponding outcomes and spatial errors are presented in Table \ref{tab:triang}. For individual observations, the spatial error arises from several factors, including angular resolution, calibration accuracy, and precision in locating the fireball trajectory.}

\begin{table*}
\caption{Observed directions and parameters for fireball FN200907. Table shows the location and observational data for stations, including the beginning and end points, along with their corresponding \textdelta z values.}
\begin{tabular}{ c c | c | c c c }
\hline
\multicolumn{1}{p{2cm}}{\begin{center}Station \end{center}} & 
\multicolumn{1}{p{2cm}}{\begin{center}Station coordinates\end{center}} & 
\multicolumn{1}{p{2cm}}{\begin{center}Point \end{center}} &
\multicolumn{1}{p{2cm}}{\begin{center}Azimuth \end{center}} & 
\multicolumn{1}{p{2cm}}{\begin{center}Altitude \newline (corrected)\end{center}} & 
\multicolumn{1}{p{2cm}}{\begin{center}\textdelta z\end{center}} \\
 & [\textdegree, \textdegree, m] &  & [\textdegree] & [\textdegree] & [m]\\
\hline
Tampere & 61.5118, 23.7931, 171m  & Begin & 346.462 & 5.501 & 197 \\
   		&& End & 11.913 & 4.149 & 300 \\
\hline
Tohmajärvi & 62.2843, 30.0679, 89m & Begin & 321.945 & 4.380 & 263 \\
  		& & End & 328.916 & 4.890 & 225 \\
\hline
Haukipudas &65.1294, 25.2905, 9m & Begin & 312.396 & 16.193 & 29 \\
   		&& End & 50.762 & 56.517 & 3 \\
\hline
\end{tabular}
\label{tab:setup}
\end{table*}

\begin{table*}
\caption{Triangulations of FN200907 against the Haukipudas observation. Table shows the triangulated beginning and ending points of the fireball for both with and without \textdelta z correction, allowing for easy comparison. These values are calculated for Tampere-Haukipudas and Tohmajärvi-Haukipudas combinations.}
\begin{tabular}{ c c c c c c c c}
\hline
\multicolumn{1}{p{2cm}}{\begin{center}Station paired \newline with Haukipudas \end{center}} & 
\multicolumn{1}{p{1.5cm}}{\begin{center}Point\end{center}} & 
\multicolumn{1}{p{1.5cm}}{\begin{center}Type \end{center}} &
\multicolumn{1}{p{1.5cm}}{\begin{center}Latitude \end{center}} & 
\multicolumn{1}{p{1.5cm}}{\begin{center}Longitude \end{center}} & 
\multicolumn{1}{p{1.5cm}}{\begin{center}Height \end{center}} &
\multicolumn{1}{p{1.5cm}}{\begin{center}Spatial \newline difference \end{center}} &
\multicolumn{1}{p{1.5cm}}{\begin{center}\edit{Spatial \newline error} \end{center}}\\
 &  &  & [\textdegree] & [\textdegree] & [m] & [m] & [m]\\
\hline
Tampere & Begin & with \textdelta z    & 66.753 & 20.583 & 88303 & 340 & \edit{214.55}\\
   		&       & without \textdelta z & 66.751 & 20.585 & 88066 & & \edit{214.65}\\
\hline
Tampere & End   & with \textdelta z    & 65.326 & 25.723 & 47278 & 404 & \edit{153.48}\\
   		&       & without \textdelta z & 65.324 & 25.722 & 46945 & & \edit{153.57}\\
\hline
Tohmajärvi & Begin & with \textdelta z    & 66.708 & 20.767 & 86797 & 1018 & \edit{146.85}\\
           &       & without \textdelta z & 66.702 & 20.781 & 86368 &  & \edit{147.06}\\
\hline
Tohmajärvi & End   & with \textdelta z    & 65.424 & 25.420 & 50065 & 608 & \edit{92.02}\\
           &       & without \textdelta z & 65.420 & 25.426 & 49767 &  & \edit{92.14}\\
\hline
\end{tabular}
\label{tab:triang}
\end{table*}

\subsection{\edit{Iron meteorite fall in Sweden (FN201107b)}}

\edit{The concept of the \textdelta z-correction has previously been applied to the Ådalen iron meteorite fall (Kyrylenko et al. 2023), which included parameters describing the visible flight of the fireball and the determination of the meteoroid's pre-entry orbit.}

\edit{The case of Ådalen fireball provides a complementary example of δz effects. The observational geometry was such that the fireball was seen nearly perpendicular to the horizon. Under these conditions, the \textdelta z-correction has minimal impact on the reconstructed trajectory but significantly affects the derived velocities.}

\edit{Without applying \textdelta z, the fitted exponential velocity model  (Whipple and Jacchia 1957) yields parameters}
\begin{equation}
b = 17,745.85, \quad c = -0.33, \quad k = 2.074.
\end{equation}
\edit{After implementing the \textdelta z-correction, the model parameters become}
\begin{equation}
b = 17,703.22, \quad c = -0.39, \quad k = 2.056.
\end{equation}
\edit{The \textdelta z-corrected velocities are lower and more realistic, producing a 560~m\,s$^{-1}$ difference at the lowest observed point (t = 4.40~s from the start), based on the Larvik video observation. The entry velocity at t = 0~s, observed at a higher altitude where \textdelta z is smaller, shows only a 43~m\,s$^{-1}$ difference.}

\edit{These velocity differences are critical for evaluating the deceleration, mass, ground impact of the meteorite, and deriving the strewn field (Moilanen and Gritsevich 2021, 2022, Moilanen et al. 2026). In contrast, the derived orbit (Kyrylenko et al. 2023) is largely insensitive to \textdelta z in this case, as the entry velocity and radiant direction remain practically unaffected.}


\section{Discussion}

\edit{In a spherically stratified atmosphere, ray propagation obeys the invariant (see Appendix)
\begin{equation}
n(r)\, r \sin\theta = \mathrm{constant},
\end{equation}
where $n(r)$ is the refractive index, $r$ is the radial distance from the Earth's center, and $\theta$ is the local zenith angle. This relation implies that all rays arriving at an observing site with the same apparent elevation angle emerge from the refractive layers of the lower atmosphere along parallel straight paths once they reach sufficiently thin air.}

\edit{Consequently, the total bending caused by refraction is completely determined by the apparent direction at the observer and the atmospheric profile, and not by the distance to the fireball. Above the lower refractive layers, the ray travels essentially in a straight line, therefore replacing the refracted ray by a straight ray originating from a vertically displaced observer yields the same geometric intersection with any distant target. This vertical displacement is exactly the \textdelta z correction.}

\edit{This explains why, for a given apparent elevation, the difference between the full finite-distance ray solution and the raised-observer geometry remains small over a wide range of fireball heights, as demonstrated in Table~\ref{tab:lower}.  In this sense, the \textdelta z correction effectively removes the explicit dependence of the reconstructed trajectory on the distance to the fireball, provided that the refraction is dominated by the lower atmosphere (Smart 1949). Airborne and space-based observations (Vaubaillon et al. 2015, Coleman et al. 2023), which largely avoid these dense refractive layers, are therefore minimally affected by atmospheric refraction compared to low-elevation ground-based measurements.}

\edit{The application of the \textdelta z correction demonstrates its importance for accurately reconstructing fireball trajectories and velocities, particularly for observations made at low apparent elevations. For the FN200907 event, triangulation performed with and without the \textdelta z correction reveals spatial displacements of order one kilometre at stations where the fireball appeared only a few degrees above the horizon. In addition to these positional effects, the Ådalen iron meteorite fall illustrates that the \textdelta z correction can also significantly influence derived velocities. Such velocity differences are critical for estimating meteoroid mass (Gritsevich et al. 2017), shape change and luminous efficiency coefficients (Gritsevich \& Koschny 2011; Bouquet et al. 2014), and meteorite fall likelihood (Gritsevich et al. 2009; 2012; Turchak \& Gritsevich 2014; Moreno-Ibáñez et al. 2020; Sansom et al. 2019; Boaca et al. 2022; Peña-Asensio et al. 2023).}

\edit{The Green model provides an analytical solution for \textdelta z and is therefore suitable for correcting absolute atmospheric refraction. In practice, the problem can be divided into two components: selecting an accurate refraction model and determining whether real atmospheric conditions must be explicitly considered. As shown in Table \ref{tab:base}, the RT and Bennett refraction models yield very similar refraction values; however, when these values are used as inputs to the Green model, the resulting \textdelta z values differ. This sensitivity highlights the importance of selecting an appropriate refraction model. When the atmospheric profile is known, the RT method is recommended to achieve the most reliable results.}

\edit{If a fireball positional accuracy of 50 m is desired, the \textdelta z correction must be applied for observed apparent angles of 12° or less. For Monte Carlo strewn-field simulations, where an accuracy of approximately 300 m is sufficient (Moilanen et al. 2021), the correction becomes necessary for observed apparent angles of 4° or less. The combination of the Green model with Bennett’s refraction model yields \textdelta z values accurate to approximately 40 m for apparent angles up to 12°. In practice, the most robust long-term solution is to integrate the \textdelta z correction directly into meteor data reduction pipelines, thereby avoiding manual decision-making and enabling automatic, consistent correction for atmospheric refraction.}

\edit{Negative apparent angles may occur at elevated observing sites; however, most conventional refraction models are defined only for angles above the true horizon. Bennett’s refraction model, for example, overcompensates \textdelta z values when used with the Green model, producing errors of approximately 200 m for negative angles and up to about 150 m even for small positive angles. These limitations further motivate the use of numerical RT model.}

\edit{As shown in Table \ref{tab:lower}, \textdelta z values for objects located at lower altitudes differ only marginally from the full \textdelta z value, consistent with the fact that most atmospheric refraction occurs in the lowest atmospheric layers. For a target positional accuracy of 50 m, δz errors for fireballs above 15 km altitude can generally be neglected. This criterion encompasses the luminous flight altitudes of most meteorite-producing fireballs (Moreno-Ibáñez et al.  2015, 2017; Sansom et al. 2019; Peña-Asensio \& Gritsevich 2025). Only a small subset of exceptionally large events, such as the Ådalen iron meteorite fall (Kyrylenko et al. 2023), exhibit luminous flight below 15 km altitude and are typically observed from large distances at small apparent angles.}

\edit{Atmospheric anomalies are expected to have only a minor influence on \textdelta z values, as shown in Table \ref{tab:inv}. Nevertheless, observations at very low apparent angles, particularly below 1°, should be treated with care, and the best available atmospheric profile data should be used when possible. Variations in real atmospheric conditions have a negligible impact on refractive indices computed using the Ciddor equation, and a spherical Earth approximation is sufficient for \textdelta z calculations.}

\edit{Based on our experience with FFN data and the results presented here, the systematic application of the \textdelta z correction \newedit{can substantially improve} the accuracy of both positional and velocity estimates. This enables the reanalysis of historical fireball events with greater precision, facilitates improved triangulation and trajectory reconstruction for current fireball networks (Jenniskens and Devillepoix 2025), and provides more reliable inputs for orbital and dynamical modelling. Consequently, predictions of strewn fields and meteorite recovery efforts can be made with increased confidence. The online \textdelta z calculator provided with this publication, together with its openly available source code, offers a convenient and reproducible means of applying these corrections across different observing sites, geometries, and atmospheric conditions.}

\section{Conclusions}


The proposed \textdelta z-correction technique is applicable to meteor observations and \newedit{can significantly improve} the accuracy of fireball data analysis. Whether implemented using simplified analytical formulations or a full RT model approach, the method is well suited for determining the three-dimensional luminous flight path of fireballs. By introducing the \textdelta z correction, fireball positions can be robustly reconstructed without explicitly accounting for the finite distance to the object or its height above the Earth’s surface. These techniques can be readily integrated into existing meteor data-reduction pipelines (Peña-Asensio et al. 2021; Vida et al. 2021; Shober et al. 2026), enabling automatic correction for atmospheric refraction in low-elevation fireball observations.

The magnitude of \textdelta z depends primarily on the apparent elevation angle, observer altitude, and atmospheric structure, with a weaker dependence on observation wavelength. Simulations show that \textdelta z attains its largest values at apparent elevations below approximately five degrees, frequently exceeding several hundred meters. Near-surface thermal inversions with temperature differences of about 5 K can modify \textdelta z by several tens of meters, whereas variations in CO$_2$ concentration, relative humidity, and wavelength have a negligible effect. Differences between spherical and ellipsoidal Earth models introduce only meter-scale variations, confirming that a spherical Earth approximation is sufficient for practical \textdelta z calculations. Overall, the \textdelta z-correction proves robust across a wide range of observing conditions, with its strongest impact occurring for near-horizon fireball geometries and non-standard atmospheric profiles.

Selecting an appropriate analytical refraction model for elevated observing sites remains challenging due to the sensitivity of the Green model to small refraction errors. In contrast, the RT model provides accurate refraction estimates, including \textdelta z, for arbitrary observing geometries and atmospheric conditions without such limitations. For observing stations at elevations exceeding 1000 m, \textdelta z values differ substantially from those at sea level, making careful model selection essential when applying the Green formulation. Particular caution is required for negative apparent angles and for defining the associated input parameters, such as refractive index. In these cases, RT model offers a more reliable solution, delivering \textdelta z corrections with sufficient accuracy that other sources of uncertainty become negligible when real atmospheric data are available.

\edit{Beyond meteor studies, the \textdelta z concept may be applicable to any ground-based optical observations of nearby objects at low apparent elevation angles, where standard refraction corrections implicitly assume sources at infinite distance. This includes observations of low-Earth-orbit satellites, space debris, aircraft, high-altitude balloons, and unmanned aerial vehicles. In such cases, neglecting finite-distance refraction effects introduces systematic, distance-dependent errors in the reconstructed viewing direction. By accounting for these effects through the \textdelta z correction, the line-of-sight geometry becomes independent of object distance, enabling more accurate three-dimensional position and velocity estimates, particularly for near-horizon observations.}

\section*{Acknowledgements}

This work was supported, in part, by the Academy of Finland project no. 325806 (PlanetS). We thank all members of the Finnish Fireball Network and acknowledge Ursa Astronomical Association for the support with the Network coordination. The programme of development within Priority-2030 is acknowledged for supporting the research at UrFU. We thank Panu Lahtinen (Finnish Meteorological Institute, FMI) for his help with obtaining the actual atmospheric data for fireball events. We are grateful to Eloy Peña-Asensio (University of Alicante) and Alberto J. Castro-Tirado (Instituto de Astrof\'isica de Andaluc\'ia, IAA-CSIC) for valuable discussions, which helped us to improve this work. Finally, we honor the legacy of Esko Lyytinen, who originally initiated the concept of the \textdelta z-correction (Gritsevich et al. 2021). This study is dedicated to his memory.

\section*{Data Availability Statement}

\subsection*{Atmospheric profile, \textdelta z calculation and source code}

\edit{An interactive \textdelta z-correction calculator\newedit{, together with the source code for the JavaScript and Python implementations,} is available at https://www.ursa.fi/$\sim$visuri/dz-calculator/}

\edit{The U.S. Standard Atmosphere (USSA) model can be downloaded with appropriate parameters here: https://www.digitaldutch.com/atmoscalc/help.htm}

\subsection*{Proof of concept case FN200907}

\edit{For this study, we used observations from two locations and one video to determine the luminous flight trajectory. These observations are publicly available at:}

\edit{- Haukipudas: https://www.taivaanvahti.fi/observations/show/935290}

\edit{- Tampere: https://www.taivaanvahti.fi/observations/show/93504}

\edit{- Tohmajärvi: https://www.taivaanvahti.fi/observations/show/93511}

\subsection*{Ådalen meteorite fall}

\edit{The previously analysed iron meteorite fall case included four stations. These observations are available at:}

\edit{- Tampere: https://www.taivaanvahti.fi/observations/show/94749}

\edit{- Nyrölä: https://www.taivaanvahti.fi/observations/show/94766}

\edit{- Kyyjärvi: Available from the authors upon request}

\edit{- Larvik: https://norskmeteornettverk.no/meteor/20201107/212700/larvik/cam2/larvik-20201107212700.mp4}

\section*{References}

Abshire, J., Riris, H., Allan, G., Weaver, C., Mao, J., Sun, X., Hasselbrack, W., Kawa, S. Biraud, S.. (2010). Pulsed airborne lidar measurements of atmospheric CO2 column absorption. Tellus B, 62., 770 - 783.

Andrade, M., Docobo, J. Á., García-Guinea, J., Campo, P. P., Tapia, M., Sánchez-Muñoz, L., ... Llorca, J. (2023). The Traspena meteorite: heliocentric orbit, atmospheric trajectory, strewn field, and petrography of a new L5 ordinary chondrite. Monthly Notices of the Royal Astronomical Society, 518(3), 3850-3876.

Auer, L., Standish, E. (2000). Astronomical refraction: computational method for all zenith angles. The Astronomical Journal, 119:2472-2474.

Bennett, G. G., (1982). The calculation of astronomical refraction in marine navigation. Journal of Navigation, Vol. 35, p. 255 - 259.

Boaca, I., Gritsevich, M., Birlan, M., Nedelcu, A., Boaca, T., Colas, F. ... Vernazza, P. (2022). Characterization of the fireballs detected by all-sky cameras in Romania. The Astrophysical Journal, 936(2), 150.

Borovicka, J., Spurny, P., Keclikova, J. (1995). A new positional astrometric method for all-sky cameras. Astronomy and Astrophysics Supplement, v. 112, p. 173, 112, 173.

Bouquet, A., Baratoux, D., Vaubaillon, J., Gritsevich, M. I., Mimoun, D., Mousis, O., Bouley, S. (2014). Simulation of the capabilities of an orbiter for monitoring the entry of interplanetary matter into the terrestrial atmosphere. Planetary and Space Science, 103, 238-249.

 Castro-Tirado, A. J., Jelínek, M., Vítek, S., Kubánek, P., Trigo-Rodríguez, J. M., de Ugarte Postigo, A., Mateo Sanguino, T. J., Gomboš, Igor. (2008). A very sensitive all-sky CCD camera for continuous recording of the night sky. Proceedings of the SPIE, 7019, 70191V

Ceplecha, Z. (1987). Geometric, dynamic, orbital and photometric data on meteoroids from photographic fireball networks. Astronomical Institutes of Czechoslovakia, Bulletin, vol. 38, July 1987, p. 222-234., 38, 222-234.

Ciddor, P. E. (1996). Refractive index of air: new equations for the visible and near infrared. Appl. Optics 35, 1566-1573 
 
Coleman, A., Eser, J., Mayotte, E., Sarazin, F., Schröder, F. G., Soldin, D., Venters, T. M., Aloisio, R., Alvarez-Muñiz, J., Alves Batista, R., Bergman, D., Bertaina, M., Caccianiga, L., Deligny, O., Dembinski, H. P., Denton, P. B., di Matteo, A., Globus, N., Glombitza, J., … Zotov, M. (2023). Ultra high energy cosmic rays The intersection of the Cosmic and Energy Frontiers. Astroparticle Physics, 147. https://doi.org/10.1016/j.astropartphys.2022.102794

Dmitriev, V., Lupovka, V., Gritsevich, M. (2015). Orbit determination based on meteor observations using numerical integration of equations of motion. Planetary and Space Science, 117, 223-235.

Egal, A., Gural, P. S., Vaubaillon, J., Colas, F.,  Thuillot, W. (2017). The challenge associated with the robust computation of meteor velocities from video and photographic records. Icarus, 294, 43-57.

Gardiol, D., Barghini, D., Buzzoni, A., Carbognani, A., Di Carlo, M., Di Martino, M., ... Zollo, A. (2021). Cavezzo, the first Italian meteorite recovered by the PRISMA fireball network. Orbit, trajectory, and strewn-field. Monthly Notices of the Royal Astronomical Society, 501(1), 1215-1227.

Green, R. M. (1985). Spherical Astronomy. Cambridge University Press.

Gritsevich, M. I. (2007). Approximation of the observed motion of bolides by the analytical solution of the equations of meteor physics. Solar System Research, 41, 509-514.

56.	Gritsevich M. I., 2008. The Pribram, Lost City, Innisfree, and Neuschwanstein Falls: An analysis of the Atmospheric Trajectories, Solar System Research, 42(5), 372-390. http://dx.doi.org/10.1134/S003809460805002X

Gritsevich, M. I. (2009). Determination of parameters of meteor bodies based on flight observational data. Advances in Space Research, 44(3), 323-334.

Gritsevich M. I., Stulov V.P., Turchak L.I., 2009. Classification of consequences for collisions of natural cosmic bodies with the Earth, Doklady Physics, 54(11), 499-503, http://dx.doi.org/10.1134/S1028335809110068

Gritsevich, M. I., Stulov, V. P., Turchak, L. I. (2012). Consequences of collisions of natural cosmic bodies with the Earth’s atmosphere and surface. Cosmic Research, 50, 56-64.

Gritsevich, M., Dmitriev, V., Vinnikov, V., Kuznetsova, D., Lupovka, V., Peltoniemi, J.,  ... Pupyrev, Y. (2017). Constraining the pre-atmospheric parameters of large meteoroids: Košice, a case study. Assessment and Mitigation of Asteroid Impact Hazards: Proceedings of the 2015 Barcelona Asteroid Day. 153-183. Springer International Publishing.

Gritsevich, M., Nissinen, M., Moilanen, J., Lintinen, M., Pyykko, J., Jenniskens, P., Brower, J., Moreno-Ibanez, M., Madiedo, J. M., Rendtel, J. (2021). The miner for out-of-this-world experience. Journal of the International Meteor Organization, 49, 3, 52-63.

Gritsevich, M. I., Moilanen, J., Visuri, J., Meier, M. M., Maden, C., Oberst, J., Heinlein, D., Flohrer, J., Castro-Tirado, A. J., Delgado-García, J., Koeberl, C., Ferrière, L., Brandstätter, F., Povinec, P. P., Sýkora, I., Schweidler, F. (2024). The fireball of November 24, 1970, as the most probable source of the Ischgl meteorite. Meteoritics \& Planetary Science, 59, Issue 7, pp. 1658-1691.

Jansen‐Sturgeon, T., Sansom, E. K., Bland, P. A. (2019). Comparing analytical and numerical approaches to meteoroid orbit determination using Hayabusa telemetry. Meteoritics \& Planetary Science, 54(9), 2149-2162. 

Jenniskens, P., Devillepoix, H. A. R. (2025). Review of asteroid, meteor, and meteorite-type links.  Meteoritics \& Planetary Science, 60(4), 928-973. 

Kyrylenko, I., Golubov, O., Slyusarev, I., Visuri, J., Gritsevich, M., Krugly, Y. N., ... Shevchenko, V. G. (2023). The First Instrumentally Documented Fall of an Iron Meteorite: Orbit and Possible Origin. The Astrophysical Journal, 953(1), 20. 

Lyytinen, E., Gritsevich, M. (2016). Implications of the atmospheric density profile in the processing of fireball observations. Planetary and Space Science, 120, 35-42. 

Meier, M. M., Welten, K. C., Riebe, M. E., Caffee, M. W., Gritsevich, M., Maden, C., Busemann, H. (2017). Park Forest (L5) and the asteroidal source of shocked L chondrites. Meteoritics \& Planetary Science, 52(8), 1561-1576.

Moilanen, J., Gritsevich, M., Lyytinen, E. (2021). Determination of strewn fields for meteorite falls. Monthly Notices of the Royal Astronomical Society, 503(3), 3337-3350. 

Moilanen, J., Gritsevich, M. (2021). A Spatial Heatmap for the 7 November 2020 Iron Meteorite Fall. Annual Meeting of The Meteoritical Society 2021. 84, 6252. 

Moilanen, J., Gritsevich, M. (2022). A Jump of an Iron Meteorite. Lunar and Planetary Science Conference, 53, 2933. 

Moilanen, J., Gritsevich, M., Visuri, J. (2026). The First Instrumentally Documented Fall of an Iron Meteorite: atmospheric trajectory and ground impact. The Planetary Science Journal, forthcoming

Moreno-Ibáñez, M., Gritsevich, M., Trigo-Rodriguez, J. M. (2015). New methodology to determine the terminal height of a fireball. Icarus, 250, 544-552.  

Moreno-Ibáñez M., Gritsevich M., Trigo-Rodríguez J.M. (2017): Measuring the terminal heights of bolides to understand the atmospheric flight of large asteroidal fragments // In the book "Assessment and Mitigation of Asteroid Impact Hazards", Springer International Publishing, pp. 129-152.

Moreno-Ibáñez, M., Gritsevich, M., Trigo-Rodríguez, J. M., Silber, E. A. (2020). Physically based alternative to the PE criterion for meteoroids. Monthly Notices of the Royal Astronomical Society, 494(1), 316-324. 

National Aeronautics and Space Administration. (1976). U.S. Standard Atmosphere. NASA-TM-X-74335. 
 
Peña-Asensio, E., Trigo-Rodríguez, J. M., Gritsevich, M., Rimola, A. (2021). Accurate 3D fireball trajectory and orbit calculation using the 3D-FIRETOC automatic Python code. Monthly Notices of the Royal Astronomical Society, 504(4), 4829-4840.  

Peña-Asensio, E., Trigo-Rodríguez, J. M., Rimola, A., Corretgé-Gilart, M., Koschny, D. (2023). Identifying meteorite droppers among the population of bright ‘sporadic’ bolides imaged by the Spanish Meteor Network during the spring of 2022. Monthly Notices of the Royal Astronomical Society, 520(4), 5173-5182.

 Peña-Asensio, E., Visuri, J., Trigo-Rodríguez, J. M., Socas-Navarro, H., Gritsevich, M., Siljama, M., Rimola, A., (2024). Oort cloud perturbations as a source of hyperbolic Earth impactors. Icarus, 408, article id. 115844.

Peña‐Asensio, E., Gritsevich, M. (2025). Inferring fireball velocity profiles and characteristic parameters of meteoroids from incomplete data sets. Journal of Geophysical Research: Planets, 130, e2024JE008382.

Peña-Asensio, E., Grèbol-Tomàs, P., Trigo-Rodrıguez, J.M., Gritsevich, M, Rimola, A., Corretge-Gilart, M., Guasch, C., Alcaraz, C., Ibañez, V., Gomez A., Gomez, J. (2025a). Unsynchronized fireball analysis and challenges in dimensionless atmospheric flight parametrization: the SPMN230522 superbolide as a case study. Revista Mexicana de Astronomía y Astrofísica Serie de Conferencias (RMxAC), 59, 87–91.

Peña-Asensio, E., Trigo-Rodrıguez, J. M., Gritsevich, M., Socas-Navarro, H., Visuri, J., Rimola, A. (2025b). Interstellar visitors and elusive extrasolar meteorites. Revista Mexicana de Astronomía y Astrofísica Serie de Conferencias (RMxAC), 59, 101–108.

Saemundsson, T. (1986). Atmospheric Refraction. Sky and Telescope, 72, 70.

Sansom, E. K., Bland, P., Paxman, J., Towner, M. (2015). A novel approach to fireball modeling: The observable and the calculated. Meteoritics \& Planetary Science, 50(8), 1423-1435.

Sansom, E. K., Rutten, M. G., Bland, P. A. (2017). Analysing meteoroid flights using particle filters. The Astronomical Journal, 153, 2. 

Sansom, E. K., Gritsevich, M., Devillepoix, H. A., Jansen-Sturgeon, T., Shober, P., Bland, P. A., ... Hartig, B. A. (2019). Determining fireball fates using the α–β criterion. The Astrophysical Journal, 885(2), 115. 

Sansom, E. K., Gritsevich, M., Devillepoix, H. A. R., Towner, M. C. (2021). An Interactive Quick-Look Tool for Fireballs and Their Initial Velocities. In 84th Annual Meeting of the Meteoritical Society, 84, 2609, 6199.

Shober, P. M.; Vaubaillon, J.; Anghel, S.; Devillepoix, H. A. R.; Hlobik, F.; Matlovič, P.; Tóth, J.; Vida, D.; Sansom, E. K.; Jansen-Sturgeon, T.; Colas, F.; Malgoyre, A.; Kornoš, L.; Ďuriš, F.; Pazderová, V.; Bouley, S.; Zanda, B.; Vernazza, P. (2026). Comparing the data-reduction pipelines of FRIPON, DFN, WMPL and AMOS: Case study of the Geminids. Astronomy and Astrophysics, Volume 705, id.A65. 

Smart, W. A. (1949). Spherical Astronomy. Cambridge University Press.

Stone, R. (1996). An accurate method for computing atmospheric refraction. Publications of the Astronomical Society of the Pacific, 108:1051-1058.

Tausworthe, R. C. (2005). Computing the Apparent Elevation of a Near-Earth Spacecraft at Low Elevation Angles for an Arbitrary Refraction Model. The Interplanetary Network Progress Report. 42-162. p. 1-10 

Trigo-Rodríguez, J. M., Lyytinen, E., Gritsevich, M., Moreno-Ibáñez, M., Bottke, W. F., Williams, I., ... Grokhovsky, V. (2015). Orbit and dynamic origin of the recently recovered Annama's H5 chondrite. Monthly Notices of the Royal Astronomical Society, 449(2), 2119-2127. 

Towner, M. C., Jansen-Sturgeon, T., Cupak, M., Sansom, E. K., Devillepoix, H. A. R., Bland, P. A., Howie, R. M., Paxman, J. P., Benedix, G. K., Hartig, B. A. D. (2022) Dark-flight Estimates of Meteorite Fall Positions: Issues and a Case Study Using the Murrili Meteorite Fall. Planetary Science Journal, Volume 3, Number 2.
 
Turchak, L. I., Gritsevich, M. I. (2014). Meteoroids interaction with the Earth atmosphere. Journal of Theoretical and Applied Mechanics, 44(4), 15-28. 

Vaubaillon, J., Koten, P., Margonis, A., Toth, J., Rudawska, R., Gritsevich, M., Zender, J., McAuliffe, J., Pautet, P.-D., Jenniskens, P., Koschny, D., Colas, F., Bouley, S., Maquet, L., Leroy, A., Lecacheux, J., Borovicka, J., Watanabe, J., Oberst, J. (2015). The 2011 Draconids: The First European Airborne Meteor Observation Campaign. Earth, Moon and Planets, 114(3–4), 137–157. https://doi.org/10.1007/s11038-014-9455-5

Vida, D., Brown, P. G., Cambell-Brown, M. (2018). Modelling the measurement accuracy of pre-atmosphere velocities of meteoroids. Monthly Notices of the Royal Astronomical Society, 479, 4, 4307-4319. 

Vida, D., Šegon, D., Gural, P. S., Brown, P. G., McIntyre, M. J. M., Dijkema, T. J., Pavletić, L., Kukić, P., Mazur, Michael J., Eschman, P., Roggemans, P., Merlak, A., Zubović, D. (2021). The Global Meteor Network - Methodology and first results. Monthly Notices of the Royal Astronomical Society, 506, 4, 5046-5074.

Vinnikov, V. V., Gritsevich, M. I., Turchak, L. I. (2016). Mathematical model for estimation of meteoroid dark flight trajectory. AIP Conference Proceedings, 1773, 1.

Visuri, J., Lyytinen, E., Sievinen J., Gritsevich, M. (2020). Correcting the Atmospheric Refraction of Fireball Observations at Low Elevation Angles and Significance of the Correction. Europlanet Science Congress, Vol. 14, EPSC2020-526.

Visuri, J., Gritsevich, M. (2021). Introducing the FireOwl: Data processing software of the Finnish Fireball Network. 84th Annual Meeting of the Meteoritical Society, LPI Communications 2609: 6093.

Wetherill, G. W., ReVelle, D. O. (1981). Which fireballs are meteorites? A study of the Prairie Network photographic meteor data. Icarus, 48(2), 308-328. 

Whipple, F.L., Jacchia, L.G. (1957). Reduction Methods for Photographic Meteor Trails. Smithsonian Contributions to Astrophysics 1, 183–206.

\appendix

{\appendix
\section{Derivation of the Ray Invariant in a Spherically Stratified Atmosphere}

In a spherically symmetric atmosphere, the refractive index depends only on the radial distance from the center of the Earth,
\[
n = n(r).
\]
Due to spherical symmetry, any ray propagates in a fixed plane containing the Earth’s center.

Let the ray path be parametrized by arc length $s$, with unit tangent vector $\hat{\mathbf t}$. The ray equation in a refractive medium follows from Fermat’s principle and may be written as
\begin{equation}
\frac{d}{ds}\!\left(n\,\hat{\mathbf t}\right) = \nabla n .
\label{eq:ray_eq}
\end{equation}
Since $n = n(r)$, the gradient $\nabla n$ is purely radial.

We introduce polar coordinates $(r,\phi)$ in the plane of propagation, and define $\theta(r)$ as the angle between the ray direction and the outward radial direction. The tangent vector can then be decomposed into radial and transverse components,
\[
\hat{\mathbf t} = \cos\theta\,\hat{\mathbf r} + \sin\theta\,\hat{\boldsymbol\phi}.
\]

Taking the transverse (azimuthal) component of Eq.~\eqref{eq:ray_eq}, and noting that $\nabla n$ has no transverse component, yields
\[
\frac{d}{ds}\!\left(n \sin\theta\right) + \frac{n \cos\theta}{r}\,\frac{dr}{ds} = 0.
\]
Using $\frac{dr}{ds} = \cos\theta$, this equation may be written as
\[
\frac{d}{ds}\!\left(n r \sin\theta\right) = 0.
\]
Therefore, the quantity
\begin{equation}
n(r)\, r \sin\theta = \text{constant along the ray}
\label{eq:ray_invariant}
\end{equation}
is conserved. This invariant represents the generalization of Snell’s law to a spherically stratified medium.


Equation~\eqref{eq:ray_invariant} provides the basis for correcting observations of sources at finite distances, such as fireballs. In practice, the \textdelta z correction is determined by finding the observer's apparent elevation for which the standard refraction path (calculated assuming a source at infinity) preserves the same ray invariant $n(r)\,r \sin\theta$ as the actual finite-distance ray. By enforcing this equality, the \textdelta z correction compensates for additional bending of the light ray at low elevations, enabling an accurate determination of the true source position despite atmospheric refraction.


\section{\edit{\textdelta z convergence using different air layer thickness}}
\edit{To validate the accuracy of \textdelta z calculations, Table \ref{tab:conv} shows convergence for different air layer thicknesses. Atmospheric height indicates the cumulative height reached from 0 m.a.s.l. during \textdelta z computation. The last column shows values computed using real atmospheric data (NOAA NCEP GFS Analysis). Increasing the layer thickness beyond 10 m provides negligible improvement in accuracy.}

\begin{table*}
\caption{\edit{Convergence of \textdelta z values for different atmospheric heights and air layer thicknesses. Atmospheric data from NOAA NCEP GFS Analysis for 7th September 2020, 18 UTC, at 65.50° N, 25.75° E.}}

\begin{tabular}{ c c c c c}
\multicolumn{5}{c}{Convergence of \textdelta z values} \\
\hline
 Atmospheric &  \multicolumn{4}{c}{Air layer thickness} \\
height& 10 m & 2 m & 0.01 m & 0.01 m \\
 & & & & (GFS data) \\\
[m] & [m] & [m] & [m] & [m]\\
\hline
10000 & 657.70 & 657.43 & 657.30 & 668.09 \\
20000 & 1041.87 & 1041.58 & 1041.49 & 1046.86 \\
30000 & 1129.86 & 1129.57 & 1129.50 & 1135.70 \\
40000 & 1148.12 & 1147.83 & 1147.75 & 1156.12 \\
50000 & 1151.88 & 1151.59 & 1151.51 & 1160.10 \\
86000 & 1153.17 & 1152.88 & 1152.81 & 1161.40 \\
\hline
\end{tabular}
\label{tab:conv}
\end{table*}


\bsp	
\label{lastpage}
\end{document}